# Effect of Scattering Efficiency in the Tip−Enhanced Raman Spectroscopic Imaging of Nanostructures in the Sub−Diffraction Limit

*A. K. Sivadasan,*[1,*,†,#] *Avinash Patsha*, [1,*,#] *Achyut Maity,*[2] *Tapas Kumar Chini,* [2] *Sandip Dhara*[1,*]

[1] Nanomaterials Characterization and Sensors Section, Surface and Nanoscience Division, Indira Gandhi Centre for Atomic Research, HBNI, Kalpakkam-603102, India

[2] Saha Institute of Nuclear Physics, HBNI, 1/AF Bidhannagar, Kolkata 700 064, India

**Abstract**

The experimental limitations in signal enhancement, and spatial resolution in spectroscopic imaging have been always a challenging task in the application of near−field spectroscopy for nanostructured materials in the sub−diffraction limit. In addition, the scattering efficiency also plays an important role in improving signal enhancement and contrast of the spectroscopic imaging of nanostructures by scattering of light. We report the effect of scattering efficiency in the Raman intensity enhancement, and contrast generation in near−field tip−enhanced Raman spectroscopic (TERS) imaging of one dimensional inorganic crystalline nanostructures of Si and AlN having large variation in polarizability change. The Raman enhancement of pure covalently bonded Si nanowire (NW) is found to be two orders of higher in magnitude for the TERS imaging, as compared to that of AlN nanotip (NT) having a higher degree of ionic bonding, suggesting the importance of scattering efficiency of the materials in TERS imaging. The strong contrast generation due to higher signal enhancement in TERS imaging of Si NW also helped in achieving the better resolved spectroscopic images than that of the AlN NT. The study provides an insight into the role of scattering efficiency in the resolution of near–field spectroscopic images.

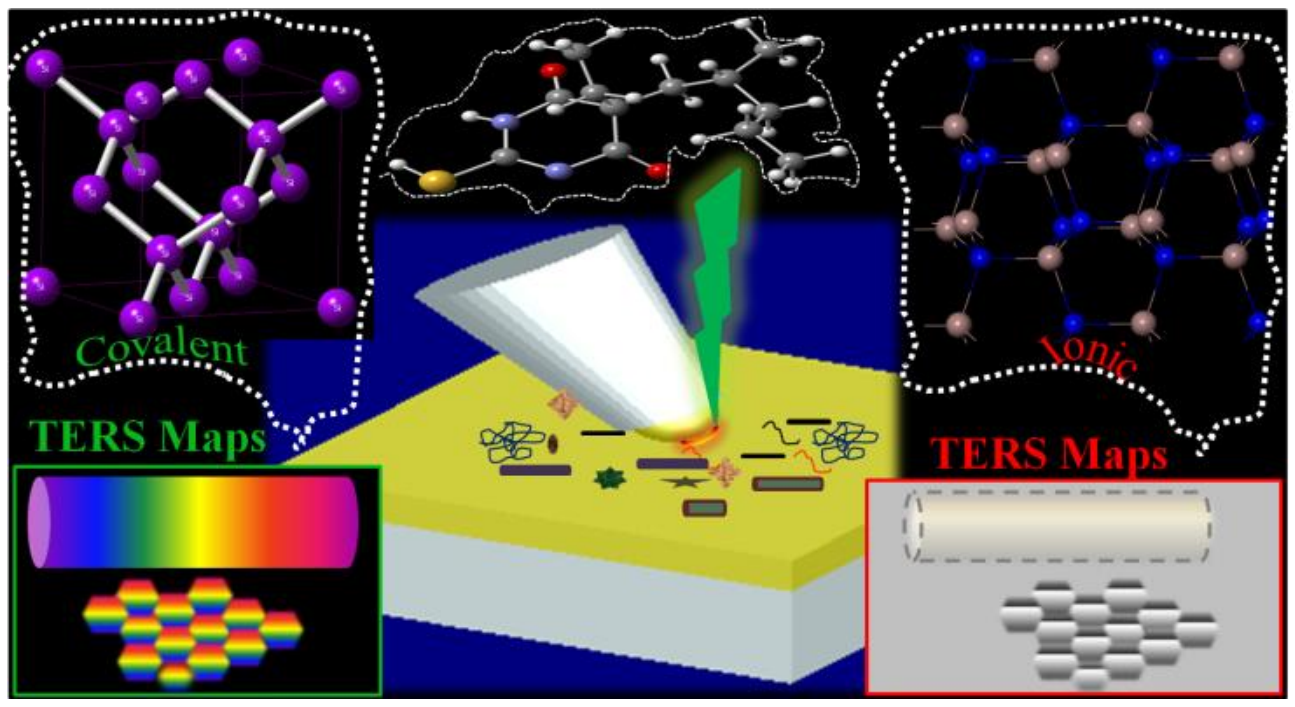

## 1. INTRODUCTION

The nanoscale focusing or imaging in the sub−diffraction limit is mainly achieved by the strong confinement or amplification of the light,[1–4] using dielectric micro–cavities and plasmonic assisted field enhancement with the help of noble metal nanostructures.[5–8] Generally, optical micro–cavities confine the electromagnetic waves in to small volumes by means of multiple reflections and sustainable standing wave formation. However, they suffer from diffraction limit. Compared to dielectric micro–cavities, nano–metallic light concentrators, resonators and metallic nanostructures are shown to the best candidates to overcome the diffraction limit of optical resolution by means of plasmonic assisted light–matter interactions.[5–8] The localized surface plasmon resonance (LSPR) of a plasmonic metal nanoparticle (NP) under the optical excitation induces strong localization of electromagnetic field near the surface of the NPs. The localized electromagnetic waves propagating along metal–dielectric interfaces in the form of surface plasmon polariton (SPP) can be guided or confined in accordance with the geometry of metallic nanostructures, for imaging of nanostructures in the sub–diffraction limit.[4,9–11]

By utilizing the plasmonic metal tip as a scanning probe, the aperture less near−field optical microscopy and spectroscopy techniques, such as scattering−type scanning near−field optical microscopy (s−SNOM), tip−enhanced Raman and photoluminescence spectroscopies (TERS, TEPL) are well explored for nanoscale imaging.[3] In case of near−field spectroscopy, TERS is the widely used technique to study different bio− and organic−molecules, carbon nanostructures, and crystalline solids with nanometer spatial resolutions.[12–15] The reported Raman enhancement factor (*EF*) for organic molecules in the near−field regime due to the presence of metal tip has the value as high as ~$10^9$ and the special resolutions (<1 nm) are obtained up to the dimension of a single molecule.[15] The *in-situ* sample preparation and TERS measurements up to single molecular level, under high vacuum ($\sim 10^{-10}$ Torr) and low temperature conditions further allowed one to observe Raman and infrared modes due to the presence of localized near−field and plasmonic electric field gradients, respectively. [16-18]

Despite the developments in the field of TERS, the Raman signal enhancement and the attainable spatial resolutions in spectral imaging are limited by the several factors such as illumination and collection geometry of the TERS setup, geometry of the tip, tip−sample distance, polarization states of the electromagnetic fields of excitation source as well as the metallic tip, energy of the incident beam, and coupling of light between the tip and sample.[19-23] Theoretical calculations show that the ultra−high spatial resolution in TERS imaging is strongly influenced by electric filed gradient effects generated around the TERS tip in the form of a cone shaped solid angle. [24] Owing to the above mentioned experimental constrains, a wide range of *EF*s from $10^3$ to $10^9$ and the spatial resolutions ranging from 30 nm to sub−nm are reported for different analytes.

Although the experimental constrains can be surmounted, the Raman scattering cross−section and the scattering efficiency, which are inherent properties of the analyte, may further limit the *EF*, and contrast variation in the image. We anticipate that the nature of chemical bonding in the analyte may play a crucial role in the Raman modes due to fluctuations in electric susceptibility with respect to the excitation source. This fact can be clearly noted from the earlier reports on different analyte systems, such as organic molecules (Triaryl dye malachite green isothiocyanate, brilliant cresyl blue, Rhodamine 6G, Thiophenol) or on carbon based nanostructures (carbon nanotubes, graphene), which are purely covalently bonded systems having high scattering efficiency.[9−11,15,22,23] The TERS studies on such analytes resulted in highest possible *EF*s ($\sim 10^{13}$) and helped for the imaging of single molecule or nanostructures in the sub−wavelength regime, especially for organic materials with a reasonable contrast. On the other hand, the studies on inorganic crystalline solids with pure covalent bonding such as Si crystal (fractional ionic character; FIC ≈ 0) showed medium enhancement of the order of $10^3$–$10^4$, however, with very few reports on imaging of single nanostructure.[25-27] The group III nitride crystals (InN, GaN, InGaN) having partially covalent and ionic bonding nature (FIC ≈ 0.5) resulted in very low Raman intensity enhancement than the covalent organic molecules and even lower than that of Si,[28-30] with hardly any reports on single nanostructure TERS imaging. In fact, not a single study on Raman enhancement of AlN is reported, as the AlN is an inorganic crystalline solid possessing highest iconicity (FIC ≈ 0.72) among other group III nitrides. The FIC for pure ionic crystals is 1.

Here we report the TERS investigations on two important inorganic nanostructures, Si nanowire (NW) and AlN nanotip (NT) having potential optoelectronic applications, to understand the influence of light scattering efficiency on TERS imaging. The present study successfully demonstrates that the Raman scattering efficiency of the analyte is also an important parameter to be considered along with other experimental parameters, for spectroscopic imaging of inorganic nanostructures using TERS technique. Unlike the earlier TERS studies on covalently bonded organic or carbon materials with large poalrizability change, the inorganic nanostructures of Si NW with pure covalent bonding (FIC ≈ 0) and AlN NT with higher FIC (~72%), and thus having different strengths of chemical bonding, are chosen as analytes in the present investigations. To the best of our knowledge, for the first time, the present study experimentally demonstrate the effect of scattering efficiency for covalently and ionic bonded inorganic semiconductor nanostructures in terms of far–field and near–field (TERS) spectroscopic imaging.

## 2. METHODS

### 2.1. Far–Field Spectral Imaging of Nanostructures: Micro–Raman Spectroscopy

The semiconductor nanostructures, used for the present studies, were transferred from the as–grown sample to high quality fused silica cover glass slips having thickness (~0.15 mm) and coated with 150 nm Au film. The far–field imaging and single spot acquisition of Raman spectra for Si NW as well as AlN NT were studied using the Raman spectrometer (inVia, Renishaw, UK) with an excitation source of 514.5 nm $Ar^+$ laser and 3000 gr·mm$^{-1}$ grating used as a monochromator for the scattered waves. A thermoelectrically cooled charged couple device (CCD) detector was used in the backscattering geometry for analyzing the Raman scattered signals. The spectra were collected using a 100× objective with numerical aperture (N.A.) value of 0.85. The corresponding focused spot size of laser beam ($d$=1.22λ/N.A.) for an excitation wavelength (λ) of 514.5 nm could be considered as ~740 nm. The fully automated motorized sample stage, having a maximum spatial resolution up to 100 nm, was used for the acquisitions of Raman signal from a predefined area of the sample containing the nanostructures. The Raman imaging of semiconductor nanostructures was performed by integrating peak intensities corresponding to the allowed Raman modes of Si NW and AlN NT, respectively. The exposure times of detector, used for recording each Raman spectra of Si NW and AlN NT, were 1 and 3 s, respectively with an excitation laser power of ~0.6 mW and the corresponding power density is 140 KW·cm$^{-2}$. The collected intensity distribution was essentially the peak intensity distribution of a particular wave number corresponding to the allowed Raman modes acquired over a predefined area and a grid resolution. In the present study, a total area of 15 × 3 μm$^2$ with 300 nm grid resolution was selected for the far–field Raman imaging of nanostructures.

## 2.2. Finite–Difference Time–Domain (FDTD) Numerical Simulations

The localized field enhancements around the coupled system of Au tip–analyte were simulated using commercially (Lumerical FDTD Solutions, Canada) available 3D finite–difference time–domain (FDTD) numerical simulations. The TERS tip was modeled by placing a gold (Au) sphere of diameter 100 nm located on a conical glass tip and the plasmon probe was simulated by confining light at the Au NP by 514.5 nm excitation wavelength. The perfectly matched–layer (PML) boundary conditions were used in the simulation. The coupled Au tip and nanostructure system were excited by a plane wave source with injection along the Z–axis direction with a polarization in the X–axis direction. The simulation time was adjusted to 700 fs so that the energy field decays fully. The mesh size was set as 2 × 2 × 1 nm during all the numerical simulations. Additionally, we used another mesh with 1.5 nm along the Z direction in the gap region between the TERS tip and the nanosructures for better accuracy. The near–field intensity maps for both analytes of Si and AlN nanostructures were extracted and compared with experimental data.

## 2.3. Near–Field Spectral Imaging of Nanostructures: Tip–Enhanced Raman Spectroscopy

TERS measurements were carried out using the scanning probe microscope (SPM) (Nanonics, MultiView 4000) coupled with the same laser Raman spectrometer (inVia, Renishaw) in the backscattering configuration. The Raman signals were collected by exciting the samples with 514.5 nm excitation and were dispersed and detected with 3000 gr·mm$^{-1}$ grating and the same thermoelectric cooled CCD detector as in the previous case. A schematic for the experimental setup of TERS is available in the figure S1of SI. The spectra were collected using a 50× objective with N.A. value of 0.42 ($d$~1.5 μm) and a laser power of ~0.6 mW with a power density of 34 kW·cm$^{-2}$. The exposure times of detector, used for recording each Raman spectra of Si NW and AlN NT, were 1 and 5 s, respectively. In order to reduce the effect of far–field artifacts and to get the improved contrast in the TERS image, the laser power and exposure times are optimized to above mentioned values. The aperture less TERS tip, used for the experiments, was an atomic force microscopic (AFM) bent glass probe working in the tuning fork feedback mechanism and was attached with an Au NP (diameter <100 nm) at the tip (Figure S1 in SI). The changes in natural frequency of the tuning fork with respect to the tip and sample interactions are considered as a feedback parameter to map the topography of the nanostructures. The tapping mode was used to operate the TERS probe and brought it into the near–field regime. In this technique, the high intense confined light at the apex of Au NP attached to the TERS tip interacted with the nanostructures and probed the enhanced Raman spectra. The 3D flat scanner, embedded at the bottom of sample stage with resolution of up to one atomic step level and with maximum vertical and horizontal displacement of 100 μm for sample scanning in the system, was used for recording the images of nanostructures. The TERS images with 256 × 256 pixels size were obtained by the integrating the peak intensity of the Raman allowed modes of nanostructures, collected from the predefined sample. The TERS setup used in the current studies provides simultaneous recording of the topography and Raman spectral information of nanostructures with high spatial resolution in the sub–wavelength regime.

## 3. RESULTS AND DISCUSSIONS

### 3.1. Morphological Analysis of Nanostructures

The morphological shape, size and distribution of Si NWs and AlN NTs are shown in the figure 1 using field emission electron microscopy (FESEM; AURIGA, Zeiss). The low magnification image of Si NWs shows cylindrical shape with very smooth surface morphology and an average length of ~10 μm (Figure 1a).

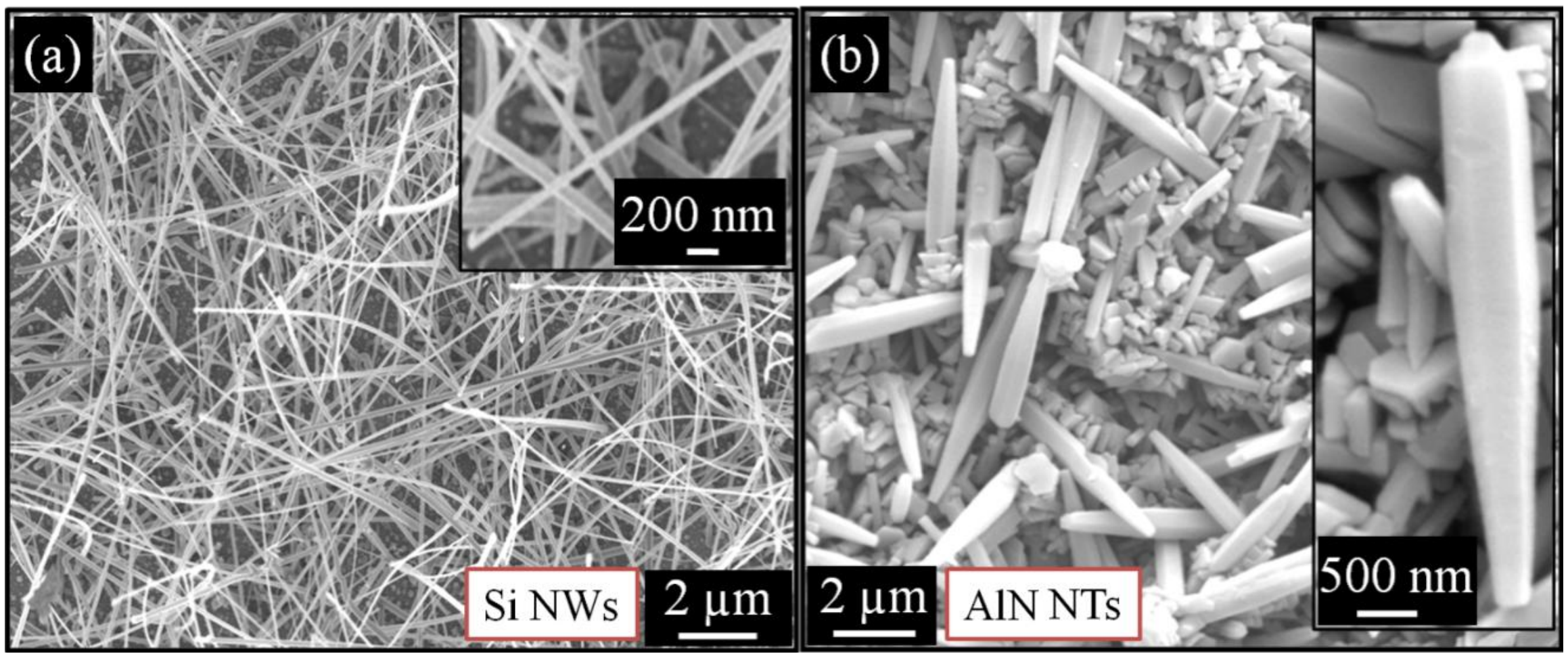

**Figure 1**. The FESEM images for the nanostructures of (a) Si NWs and (b) AlN NTs. The insets of both figures show the magnified images of nanostructures.

The inset figure 1a shows the high magnification image of Si NWs, which shows the different size variation of the NWs. The diameter variation of 20–60 nm is observed. Whereas, a uniform sized AlN NTs with hexagonal cone shape and an average length of ~5 μm is observed in the low magnification image (Figure 1b). The gradual increase in diameter (200-600 nm) of AlN NT is observed from tip to base region as shown in the inset figure 1b.

## 3.2. Far–Field Spectral Imaging of Nanostructures: Micro–Raman Spectroscopy

### 3.2.1. *Single spot Raman spectra of Si NW and AlN NT*

The Raman allowed vibrational modes (Figure 2) for a single Si NW as well as AlN NT were studied using conventional far–field micro–Raman spectrometer. The observed sharp and intense peak around ~518 $cm^{-1}$ for Si NWs (Figure 2a) is assigned as the characteristic TO phonon mode of Si. The red–shift (~2.5 $cm^{-1}$) in the peak position of Si NW from that of bulk Si (520.5 $cm^{-1}$), might be due to the light–induced non-thermal population of optical phonons in NW by high optical excitation intensity (140 KW·$cm^{-2}$).[31] In the spectra of AlN NT sample, the Raman modes centered at 610, 657, 669, 895, and 910 $cm^{-1}$ (Figure 2b) are assigned as the $A_1$(TO), $E_2^{H}$, $E_1$(TO), $A_1$(LO) and $E_1$(LO) modes of wurtzite AlN, respectively.[32,33] The Lorentzian fitted high intense peak of $A_1$(LO) mode for the AlN wurtzite NT is shown in the inset figure 2b. The sharp and high intense Raman peaks, observed for both the samples, assures the high quality crystalline nature of the semiconductor nanostructures. Moreover, the intensity of allowed Raman mode of covalent Si NW is found to be higher by one order compared to that of AlN.

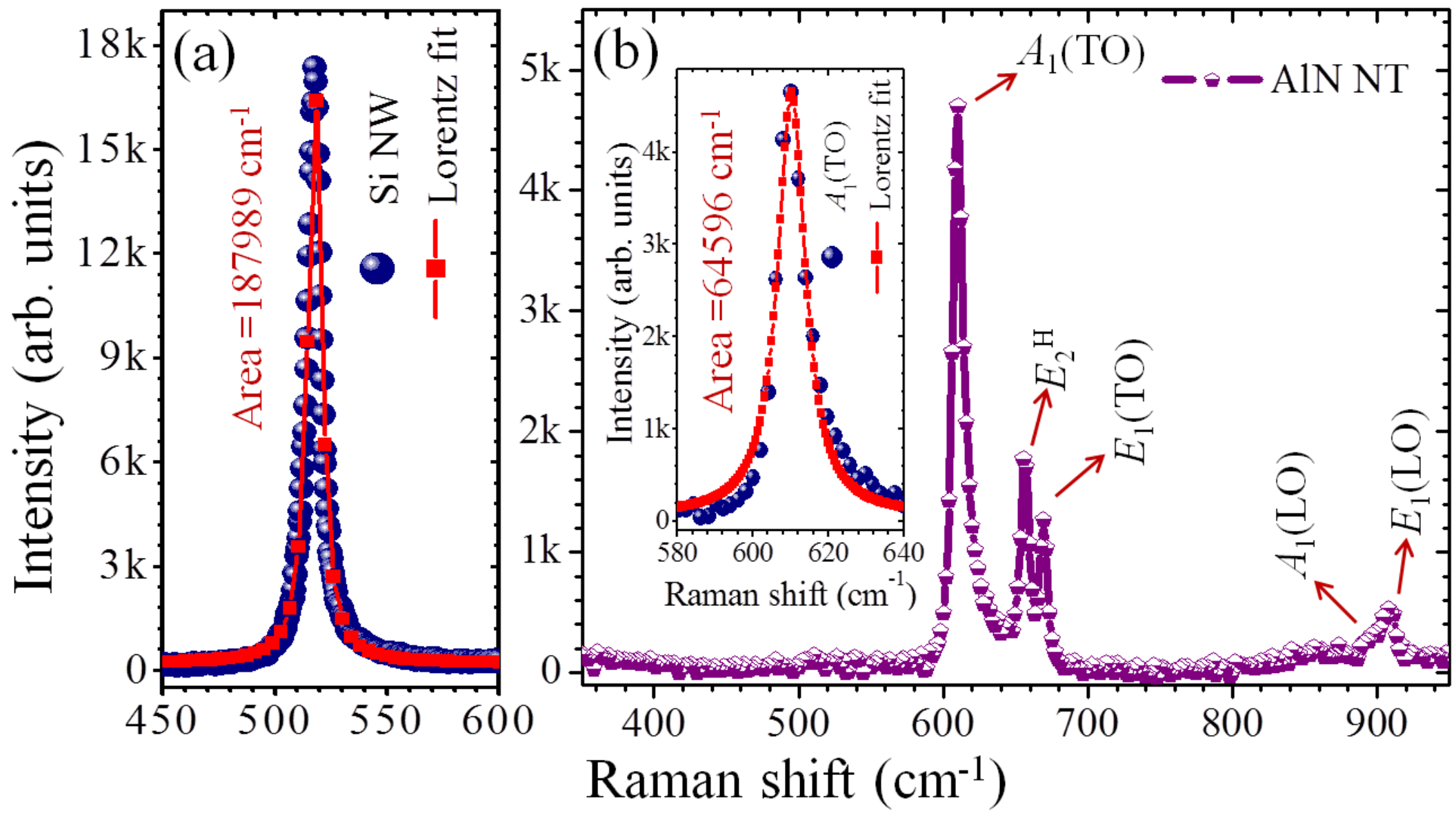

**Figure 2.** The single spot far–field Raman spectra of semiconductor nanostructures (a) Si NWs (b) AlN NT; the inset figure shows the Lorentzian fitted $A_1$(TO) mode.

The Raman scattering is an inelastic process which occurs as a result of change in the electric polarizability of the material induced by the excitation source of light. The change in polarizability of the material is associated with either creation or annihilation of various elementary excitations such as phonons and plasmons in the solid. It is well known that, in the case of Raman scattering by phonon, the scattering efficiency is higher in covalent crystallites compared to that of ionic crystallites.[32,34] The valence electrons in covalent crystals are less localized compared to that of ionic crystals, which may lead to the higher fluctuation of electric susceptibility and hence the higher change in polarizability of the covalent material induced by the creation or annihilation of phonons than that of ionic ones. Therefore, the covalent crystals possess a low polarizability. They, however show large change in the magnitude of polarizability with respect to the elementary excitation. Whereas, ionic crystals possess higher polarizability and lower change in the magnitude of polarizability compared to that of covalent crystals. The differential scattering cross section; $d\sigma$ for Raman scattering is equals the ratio of scattered to incident power per unit solid angle $\Omega$ subtended by the spectrometer. The differential scattering cross section related to the Raman scattering efficiency $S$ is given by,[35,36]

$$\frac{dS}{d\Omega} = \frac{1}{V}\frac{d\sigma}{d\Omega} \quad (\mathrm{m^{-1}\cdot sr^{-1}}) \qquad \text{..........................................(1)}$$

where $V$ is the effective scattering volume (Figure S1 in SI) of the solid or scatterer which can be limited by the beam spot size; $d$ of the excitation laser. In addition, the $d\sigma$ for the spontaneous light scattering from a solid is proportional to the fluctuation of electric susceptibility (change in polarizability) of the material with respect to the incident and scattered polarization of electric field *i.e.*, $\left|\hat{e}_2.\delta\vec{\chi}.\hat{e}_1\right|^2$; where $\hat{e}_1$ and $\hat{e}_2$ are the unit polarization vectors for the incident and scattered photons, which establish the condition for the Raman selection rule for the polarization configurations; $\delta\vec{\chi}$ is the matrix elements for the transition electric susceptibility from one state to another.[32,36] In the case of Raman scattering process, the transition electric susceptibility is a second ranked tensor operator; with $\mu\upsilon$ components; $\left(\delta\vec{\chi}\right)_{\mu\upsilon} = \left\langle f\left|\delta\vec{\chi}_{\mu\upsilon}\right|i\right\rangle$ with initial; $\left|i\right\rangle$ and final; $\left|f\right\rangle$ state of the crystal which determines condition on the transformation properties of the states under symmetry operations of the crystal space group.[36] The power of the Raman scattered light $P_S$ after the interaction with the material is proportional to the area; $A$ under the peak belongs to an allowed phonon mode in the Raman spectrum per acquisition time $t$,[35]

$$P_S \propto \frac{A}{t} \qquad \text{..................................................(2)}$$

With the help of Lorentzian curve fitting we have measured the integrated area under the curve of Raman mode for Si NW (Figure 2a) as well as the $A_1$(TO) (inset Figure 2b) mode for AlN NT. By using the equation (2), we have calculated the ratio of scattered power of the TO Raman mode of Si NW to the $A_1$(TO) mode of AlN; as $P_S$(Si)/$P_S$(AlN) ~12. In the usual laboratory experimental conditions, by considering the entire scattering process is isotropic in the solid angle $\Omega$ subtended by the spectrometer, the power of scattered light $P_S$ in unit solid angle $\Omega$ per unit volume $V$ is proportional to the scattering efficiency $S$ as,[35,36]

$$\frac{dS}{d\Omega} \propto \frac{1}{V}\frac{P_S}{P_L d\Omega} \quad \text{..........................................(3)}$$

where $P_L$ is the incident laser power that induces polarizability changes in the material. All the scattering parameters depicted in the equation (3) need corrections with respect to the various possible light–matter interactions other than the refraction, reflection and absorption of incident and scattered light.[35] In the present study, we can assume that the scattering volume (depending on the laser spot size; $d$ ~740 nm; Figure S1in SI) as well as solid angle $\Omega$ subtended by the spectrometer for the scattered laser is similar for both Si NW and AlN NT. The external laser power; $P_L$ used for each measurement was kept constant (~0.6 mW; power density 140 $KW{\cdot}cm^{-2}$). Therefore, it is obvious that from the equation (3), the scattering efficiency of Raman mode corresponding to the Si NW can be higher at least 12 times (or approximately one order) as compared to that of Raman modes of AlN NT. This observation quantitatively concludes that the fluctuation of electric susceptibility (change in polarizability) of the covalent Si NW (FIC $\approx$ 0) under excitation field is higher by one order compared to that of AlN NT which is having very high FIC of 0.72.

### 3.2.2. *Micro–Raman spectral mapping of Si NW and AlN NT*

The far–field micro–Raman spectroscopic images for the inorganic semiconductor Si NW as well as AlN NT is shown in the figure 3. The area selected for the Raman spectral map of Si NW and AlN NT in the optical images are shown in the figures 3a and 3b, respectively. The corresponding Raman spectral map of Si NW and $A_1$(TO) mode of AlN NT with intensity scale are showed in the figures 3c and 3d, respectively. The variation in the color scale of each map is plotted in accordance with the variation of peak intensity value of the Raman modes along the nanostructures. It is also observed that the intensity map of Raman allowed mode of Si NW is higher by one order compared to that of with AlN NT because of the higher scattering efficiency for the covalent Si crystal compared to that of with AlN crystal. Owing to the sub–wavelength sizes of Si NWs (~20-60 nm) and tip region of AlN NT (~200 nm), the optical resolution of the far–field Raman spectroscopic mapping is observed to be very poor quality.

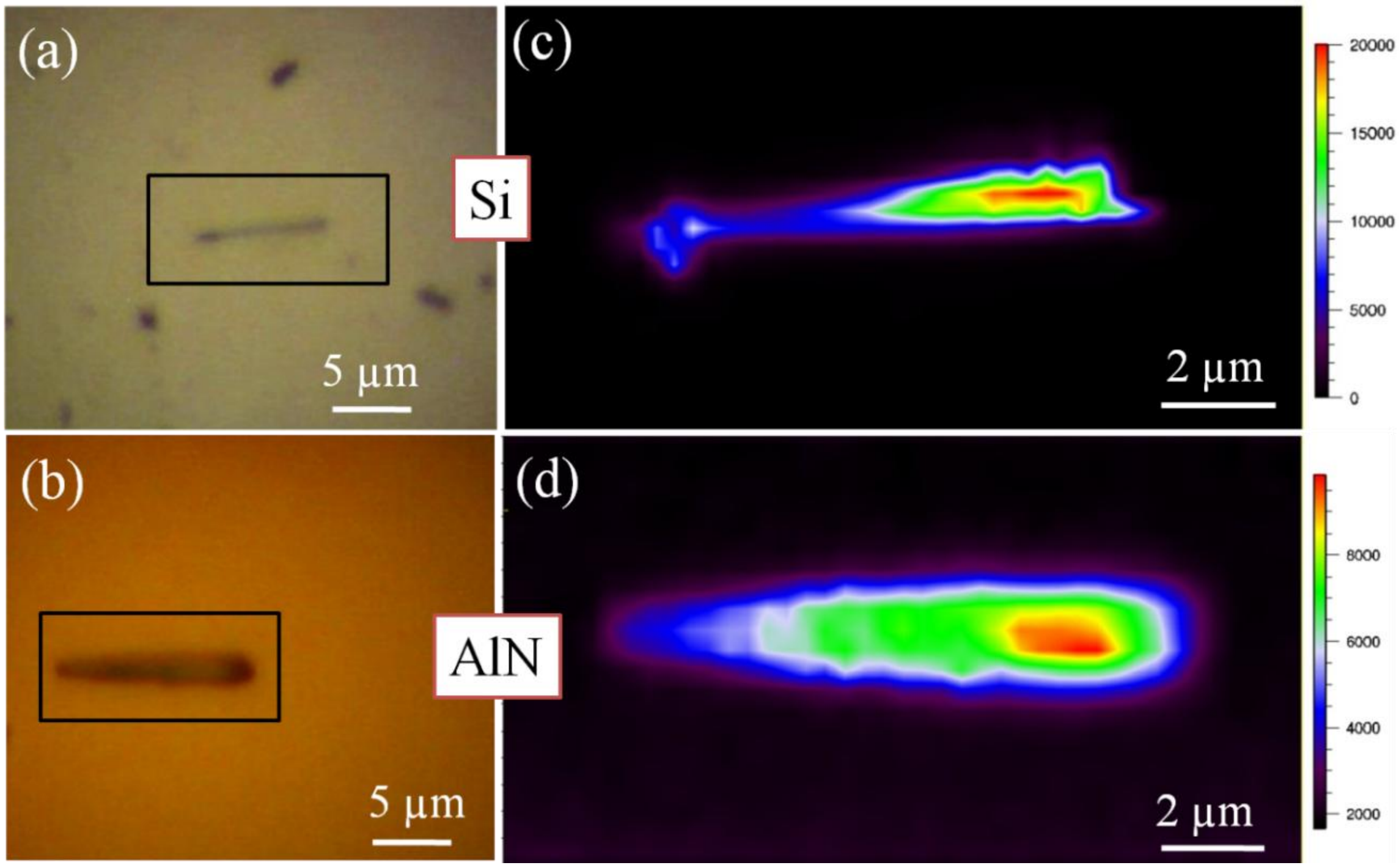

**Figure 3.** The optical image of (a) Si NW (b) AlN NT. The areas enclosed by rectangular boxes are selected for Raman imaging. The corresponding far–field micro–Raman spectroscopic mapping of (c) Si NW and (d) $A_1$(TO) mode of AlN NT. The color scale corresponding to the peak intensity value of Raman allowed modes along the nanostructures.

## 3.3. Near–Field Imaging of Nanostructures: Tip–Enhanced Raman Spectroscopy

### 3.3.1. *Topography analysis of nanostructures by AFM*

In order to study the near–field Raman spectroscopy along with spectral mapping of the inorganic semiconductor Si NW as well as AlN NT, we have selected an isolated nanostructure for each case whose dimensions are in the sub–diffraction regime, as shown in the figure 4. The 3D AFM topographic profile of the area selected for spectroscopic TERS imaging of Si NW (Figure 4a) and AlN NT (Figure 4b).

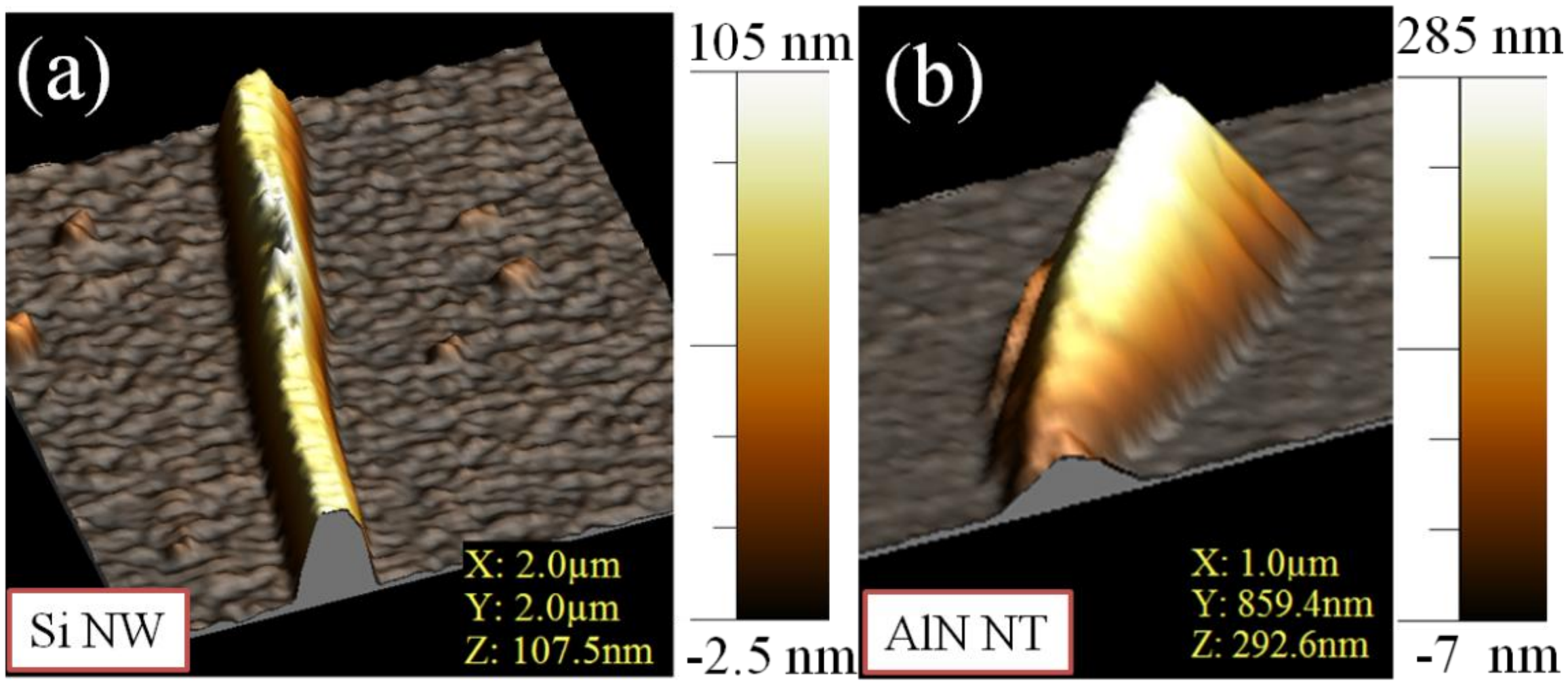

**Figure 4**. The 3D topography of (a) Si NW and (b) AlN NT; intensity scale variation corresponds to the height of the nanostructures. The height profiles of the AFM topography are available in the figure S2 of SI.

The detailed analysis of the size of these nanostructures from the height profiles in the AFM topography is available in the figure S2 of SI. The diameter of Si NW is found to be ~50-60 nm. Whereas, in the case of AlN NT, the diameter varies from ~250 nm to 70 nm from one end (base) to the other (tip). Therefore, the AFM topographic analysis ensures that the size of both the crystals with different bonding nature are far below the diffraction limit for an excitation source of 514.5 nm (~$\lambda/2$ = 257 nm). The intensity scale variation for the AFM topography for the figures indicates the height variations in the scanned area of the sample.

### 3.3.2. *Near–Field distribution maps of coupled system by FDTD simulations*

In order to study the localized field enhancement due to TERS tip on Si NW and AlN NT, the near–field intensity maps for both the systems coupled with TERS tip were simulated and extracted using FDTD numerical simulations. The schematics of the simulated structures projected in XZ–plane for Si NW (Figure 5a) and for AlN NT (Figure 5b) are shown. The near–field distribution maps projected in XZ and YZ–planes for coupled Si NW–Au, and AlN-Au systems show that the field is confined at the interface in the region of less than 50 nm spot size. The stronger electric field enhancement at the Si NW–Au NP interface (Figures 5c, 5d) as compared to AlN NT–Au NP interface (Figures 5e, 5f) is observed. The confined field strength for Si NW–Au NP interface is approximately two times that of the AlN NT–Au NP interface.

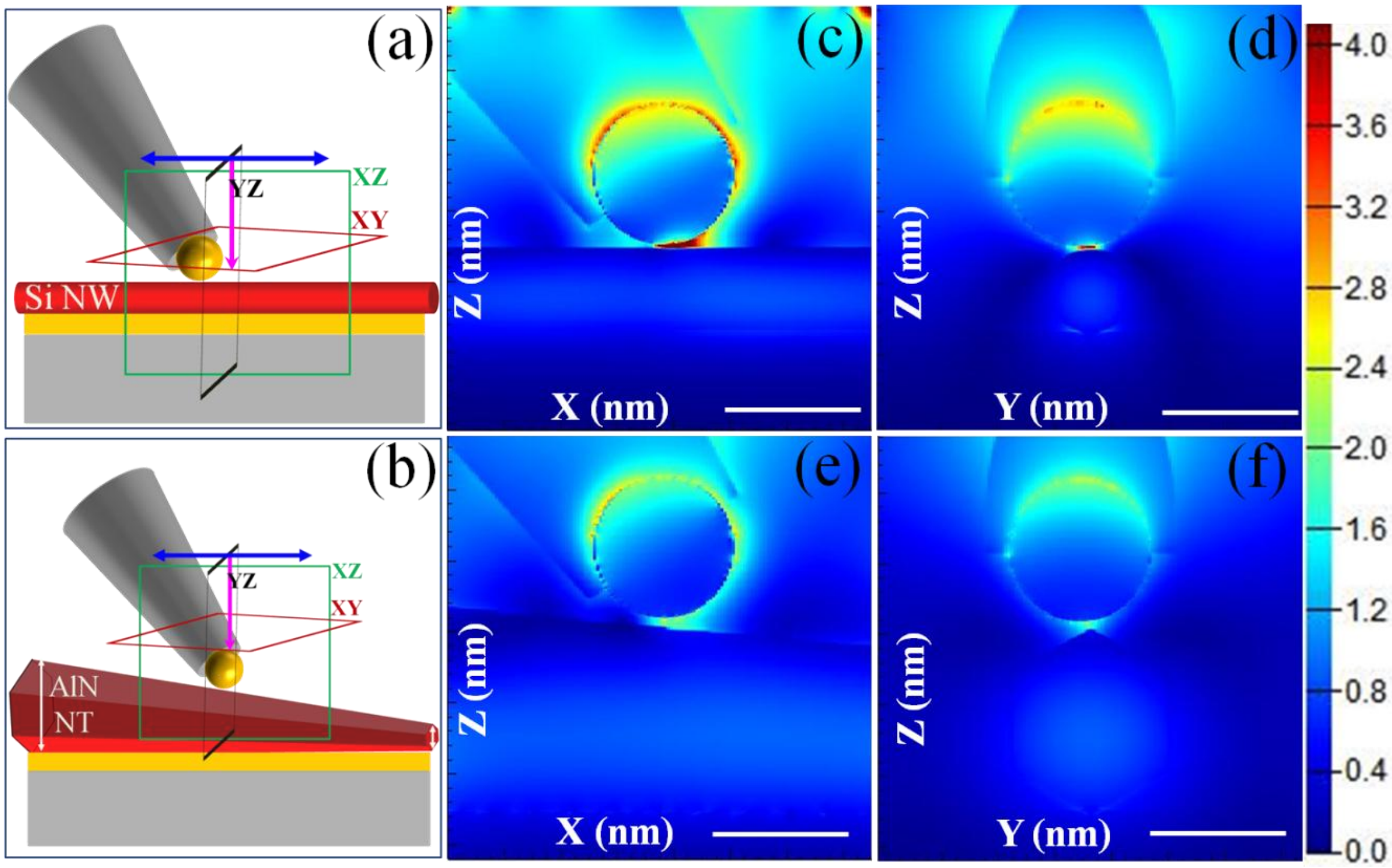

**Figure 5:** The schematics of simulated structures projected in XZ–plane for coupled TERS tip with (a) Si NW and (b) AlN NT. The near–field distribution maps in XZ and YZ–planes for Si NW (c,d) and AlN NT (e,f), respectively. Scale bar is 100 nm. The color bars indicate relative strength of the field.

### 3.3.3. *Single spot normal and tip–enhanced Raman spectra of Si NW and AlN NT*

The single spot Raman spectra of Si NW (Figure 6a) and AlN NT (Figure 6b) for far–field (without tip) and near–field (with tip) configurations are recorded. The reduction in the intensity of Raman spectra of both the nanostructures in the far–field collection compared to that of the spectra shown in the figure 2, is due to the change in the collection optics which were necessary for TERS mapping. The TERS spectra of both Si NW as well as AlN NT show a significant enhancement in the presence of plasmonic Au NP (~100 nm) attached apertureless TERS probe. The observed significant enhancement in the Raman spectra is due to the plasmonic enhancement by TERS tip and the strength of the enhancement is influenced by the localized electric field and its gradient distributions near to the probe–sample interface.[19-24] The interesting observation is that the plasmonic enhancement observed for the Raman spectrum of the Si NW (Figure 6a) even for lower acquisition time (1s) is significantly higher with respect to that of AlN NT (Figure 6b) collected with a higher acquisition time (5s).

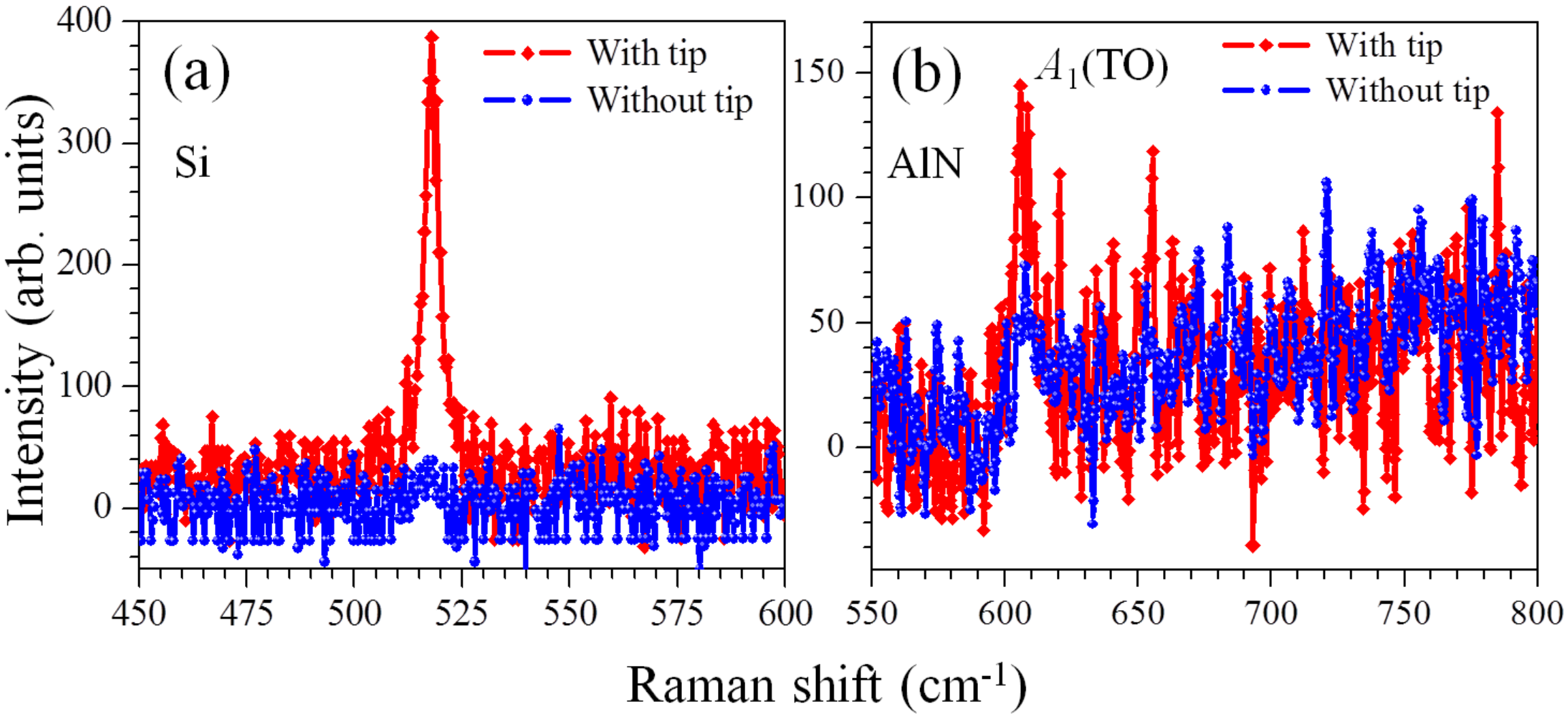

**Figure 6.** The single spot Raman spectra of semiconductor nanostructures for far–field (without tip) and near–field (with tip) configurations (a) Si NW with an acquisition time of 1s (b) AlN NT with an acquisition time of 5s.

The Raman *EF* for the collected TERS spectra is calculated using,[37]

$$EF = \left( \frac{I_{with-tip}}{I_{without-tip}} - 1 \right) \frac{V_{FF}}{V_{NF}} \qquad \text{---------------------------------------(4)}$$

where $I_{with-tip}$ and $I_{without-tip}$ are the Raman peak intensities recorded with the tip in the near– and far–field regime of the sample, respectively; $V_{FF}$ is the interaction volume of the far–field laser probe and $V_{NF}$ is the effective interaction volume of near–field TERS probe. We consider a cylindrical interaction volume for 1D nanostructure as $V = \pi(\delta/2)^2 d$ (Figure S1in SI); where, $\delta$ is the depth of penetration. In the case of Si NW, skin depth $\delta = \sqrt{\frac{\rho}{\pi . f . \mu_r}}$ is ~3 mm for a frequency of 5830.90 MHz (514.5 nm), and resistivity of ~ 2 Ωcm and $\mu_r$ ~1. Therefore, in the case of far–field Raman spectral acquisition, $V_{FF}$ for Si NW can be calculated as ~4.2×10$^{-3}$ μm$^3$ by considering $d$ ~1.5 μm and $\delta$ ~ 60 nm, which is the size of the NW diameter as the calculated skin depth (~ 3 mm) is very large with respect to it. Whereas, in the case of near–field Raman spectral acquisition, the presence of TERS probe may confine the laser light spot diameter in to ~50 nm (according to FDTD simulated near–field distribution maps). Therefore, $V_{NF}$ ~0.14×10$^{-3}$ μm$^3$ by assuming that the evanescent waves can be penetrate to the dielectric materials up to a distance of the order of a few half–wavelengths. The value of $EF_{\text{Si-NW}}$ is calculated as ~12×10$^3$. In the case of AlN NT, $V_{FF}$ can vary from the base to the tip of NT as ~ 5.8×10$^{-3}$ to 73.8×10$^{-3}$ μm$^3$ by considering $d$ ~ 1.5 μm and $\delta$ ~70 to 250 nm. Similarly, in presence of TERS probe, $V_{NF}$ may vary from ~0.19×10$^{-3}$ to 0.98×10$^{-3}$ μm$^3$. The values of $EF_{\text{AlN-NT}}$ are estimated and found to show variations from ~30 to 74. It is very clear that from the ratio $EF_{\text{Si-NW}}/EF_{\text{AlN-NT}}$ ~400, the Raman enhancement due to TERS tip is two orders of higher magnitude in case of Si NW as compared to that of AlN NT, for the similar experimental conditions. However, the FDTD simulated near–field distribution maps show that the enhanced field strength at the interface of TERS probe–sample is of the same order for Si and AlN nanostructures. The large variation in *EF* of two nanostructures suggests the role of Raman scattering cross-section (scattering efficiency), which is an intrinsic property of the material, in the Raman scattering process under near–field excitation.

### 3.3.4. *Tip–enhanced Raman spectral mapping of Si NW and AlN NT in the sub-diffraction limit using an apertureless SPM probe*

The apertureless scattering–type TERS tip, which is a bent glass probe attached with an Au NP (diameter $\leq$100 nm) at the apex of the probe is used for simultaneously recording the topography and TERS maps. The simultaneous near–field TERS and topographic imaging of Si NW (Figures 7a and 7b, respectively) and AlN NT (Figures 7c and 7d, respectively) were performed, with similar experimental conditions used for the single spot spectral acquisitions. The TERS images and the corresponding 2D topography maps of Si NW and AlN NT are compared for the dimensions of nanostructures. The color bar in the TERS map corresponds to the variation in intensity of Raman allowed mode along the nanostructures. Owing to the high signal enhancement in the Raman mode of Si by TERS tip, the strong contrast generation is observed along the NW in the TERS map of Si NW (Figure 7a). The TERS map of Si NW (Figure 7a) showed that the size of NW is exactly matched with that of topography map (Figure 7b). Therefore, in case of Si NW the TERS clearly resolved the dimension of NW, which is far below the diffraction limit for an excitation source of 514.5 nm. At the bottom portion of the Si NW, the intensity profile of TERS map is observed as higher compared to that of the top of Si NW.

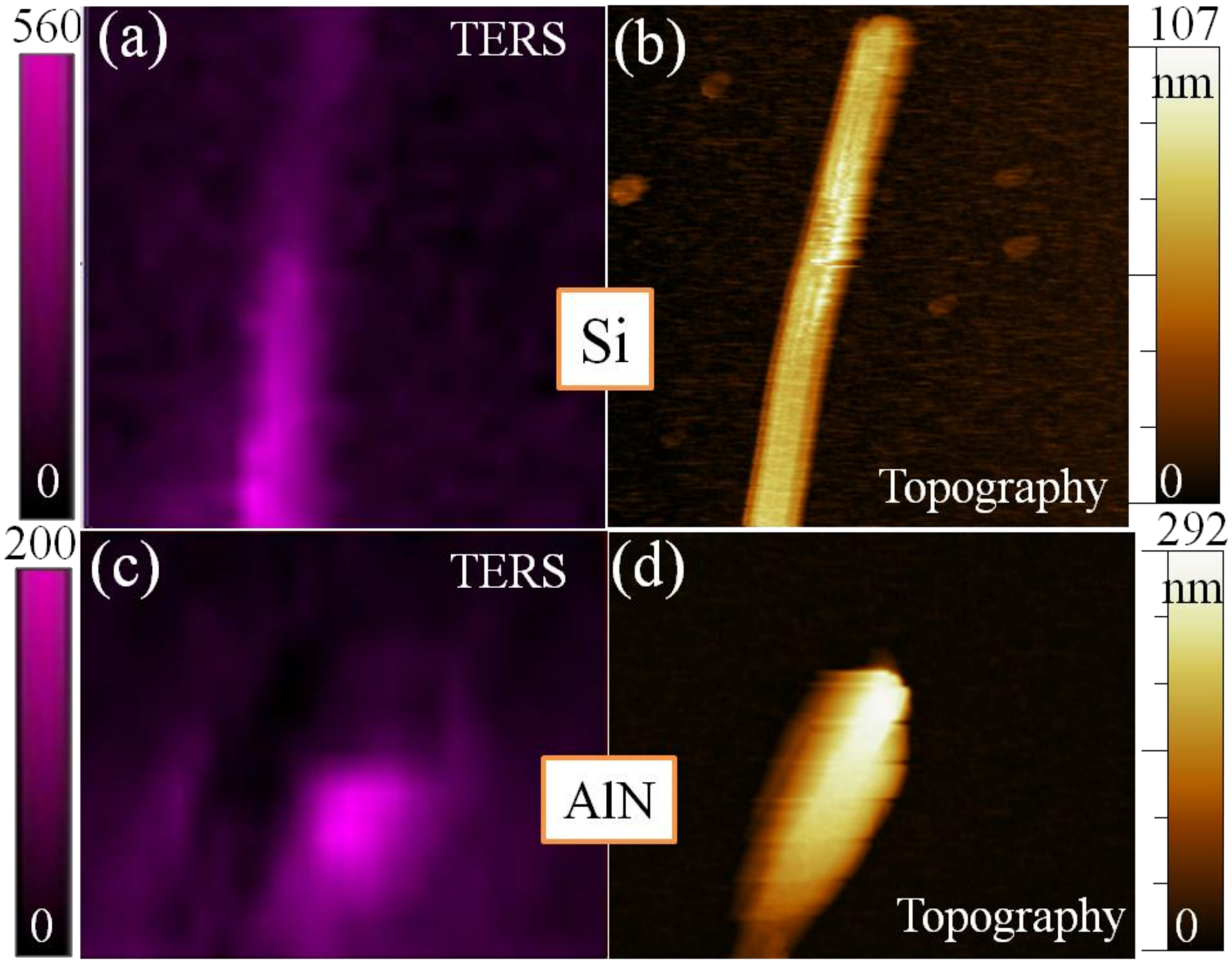

**Figure 7.** The near–field (a) TERS and corresponding (b) topography map of Si NW; (c) TERS and corresponding (d) topography maps of AlN NT. The color scale bars indicate the variation in intensity of Raman allowed mode (left) and height (right) along the NW.

The dimension of Si NW (~50-60 nm) in TERS map is exactly matching with the AFM height profile and topography maps (Figures 4a, 7b and S2 in SI). Whereas, in the case of AlN NT (Figures 7c, 7d), the resolution of the TERS image is deteriorated by the poor spectral contrast differences along the nanostructure. Even though, the nanostructure size (70-250 nm) as well as the acquisition time of 5s, used for the recording of $A_1$(TO) mode of AlN NT, are higher compared to that of Si NW (size ~50-60 nm; acquisition time of 1s), the contrast and resolution of the TERS map is better for the Si NW with lower dimension. While performing the TERS studies of nanostructures in the present investigations, it is ensured that the major contribution for scattered signal is originated mainly by near–field interaction, which is achieved by adjusting the acquisition time for recording the Raman spectra such a way that the vibrational modes are observable only in the presence of TERS probe at the vicinity of the sample. In the case of near–field single spot Raman spectrum of Si NW (Figure 6a), we can see that the major intensity of the vibrational mode generated only with the presence of TERS probe at the vicinity of the NW. Therefore, the Raman scattered signal is generated only from the interaction of material with evanescent waves, which leads to the higher contrast and thus a better resolved Raman spectral map (Figure 7a) is achieved. Similarly, we have also followed the same procedure in the case of AlN NT to get the scattered signal only from the evanescent waves by adjusting the acquisition time of 5s as mentioned in the earlier case.

From the FDTD simulations of near–field distribution maps and the experimental Raman enhancement results, it can be concluded that Raman enhancement is purely of electromagnetic origin. However, the variation in the enhancement factor is due to the influence of field strength as well as the Raman scattering efficiency of a material under excitation field, simultaneously. As a result, a strong contrast in the TERS map of Si NW is observed. We have already observed that the fluctuations in electric susceptibility of covalent Si NW are higher by one order compared to that of AlN NW. Therefore, the higher scattering efficiency of Si NW leads to the better resolved TERS map even for the lower acquisition time compared to that of AlN NT. Moreover, from the equation (1), it is very clear that the scattering efficiency is inversely proportional to the effective scattering volume; $V$ which is limited by the beam spot size; $d$ of the excitation laser. With the help of Au NP (~100 nm) attached to the TERS tip used in the present study, it is possible to confine and concentrate the excitation light beam size in to a dimension less than 100 nm. The tiny light spot at the vertex of TERS probe leads to achieve the high resolution in the spectral mapping as well as it may help to reduce $V$ into nano size regime which may help to improve the scattering efficiency, as well. Since the light is confining in to a tiny volume with respect to the excitation spot size determined by the dimensions of apertureless TERS probe, one can also expect the appearance of non–zone centre phonon modes in the spectra because of the non–conservation of wave vector; $q$ with $\Delta q \sim 2\pi/d$ as well as the damping of incident and scattered waves in the tiny scattering volume.[36] However, the plasmonic assisted confinement of light as well as the better scattering efficiency of the covalent Si NW helped us to achieve a spectroscopic image of the 1D inorganic semiconductor NW in the sub–diffraction regime, for the first time. The present study provides a physical insight into the scattering efficiency of a material under investigation and its influence on the near–field spectroscopic imaging techniques.

## 4. CONCLUSIONS

The spectroscopic imaging of inorganic crystalline nanostructures using the Raman scattering is investigated. The tip–enhanced Raman spectroscopic (TERS) imaging of covalently bonded Si nanowire (NW) and AlN nanotip (NT) with more ionic nature than covalent, were used for understanding the influence of scattering efficiency, an intrinsic property of the material in the spectroscopic imaging of inorganic nanostructures in the sub–diffraction regime. The single spot far–field Raman intensity of Si NW of diameter ~ 50-60 nm is observed to be one order higher compared to that of AlN NT (70-250 nm) because of the influence of the variation in fluctuation of electric susceptibility of two nanostructures under excitation field. Owing to the Abbe's diffraction limit for a wavelength of 514.5 nm as well as higher excitation beam diameter (~740 nm), the spatial resolution of the far–field Raman spectroscopic mapping of inorganic nanostructures are observed as very poor in quality. Whereas, the near–field Raman spectroscopic studies along with imaging of the covalent Si NW as well as partially ionic AlN NT showed the contrast variation in the TERS map. The FDTD simulated near–field distribution maps confirmed the presence of strong localized field enhancement at the interface of TERS tip and analyte (Si NW, AlN NT) in a region of 50 nm. The TERS enhancement factor of Si NW is found to be two orders higher than that of AlN for the similar experimental conditions. The Raman enhancement is found to be purely of electromagnetic origin. However, the variation in the enhancement factor is due to the simultaneous effect of field strength as well as the scattering efficiency of the material under excitation field. Eventually, the contrast and thus the resolution of the TERS map for covalent Si NW is found to be better than that of AlN NT due to the effect of variation in scattering efficiency.

**ASSOCIATED CONTENT**

**Supporting Information**

The Supporting Information is available free of charge on the ACS Publications website at DOI:

Schematic for the experimental setup for TERS

**AUTHOR INFORMATION**

Corresponding Authors

*E-mail: sivankondazhy@gmail.com.

*E-mail: avinash.phy@gmail.com.

*E-mail: dhara@igcar.gov.in.

† Presently at Government Higher Secondary School, Pazhayannur-680587, Kerala, India

**AUTHOR CONTRIBUTIONS**

#These authors contributed equally.

AKS planned the work, assisted in TERS, performed the far−field micro−Raman measurements and drafted the manuscript. AP carried out the TERS experiments, simulations, result analysis, and aided in the final drafting. SD guided the experimental proceedings and contributed in critical interpretations. All authors discussed the results, commented on the manuscript and gave approval to the final version of the manuscript.

**Notes**

The authors declare no competing financial interest.

**ACKNOWLEDGEMENTS**

Our sincere thanks to Kishore K. Madapu and Santanu Parida of SND, IGCAR, for their suggestions and useful discussions in performing the experiments. AKS and AP thank the department of atomic energy for financial support.

## Supporting Information

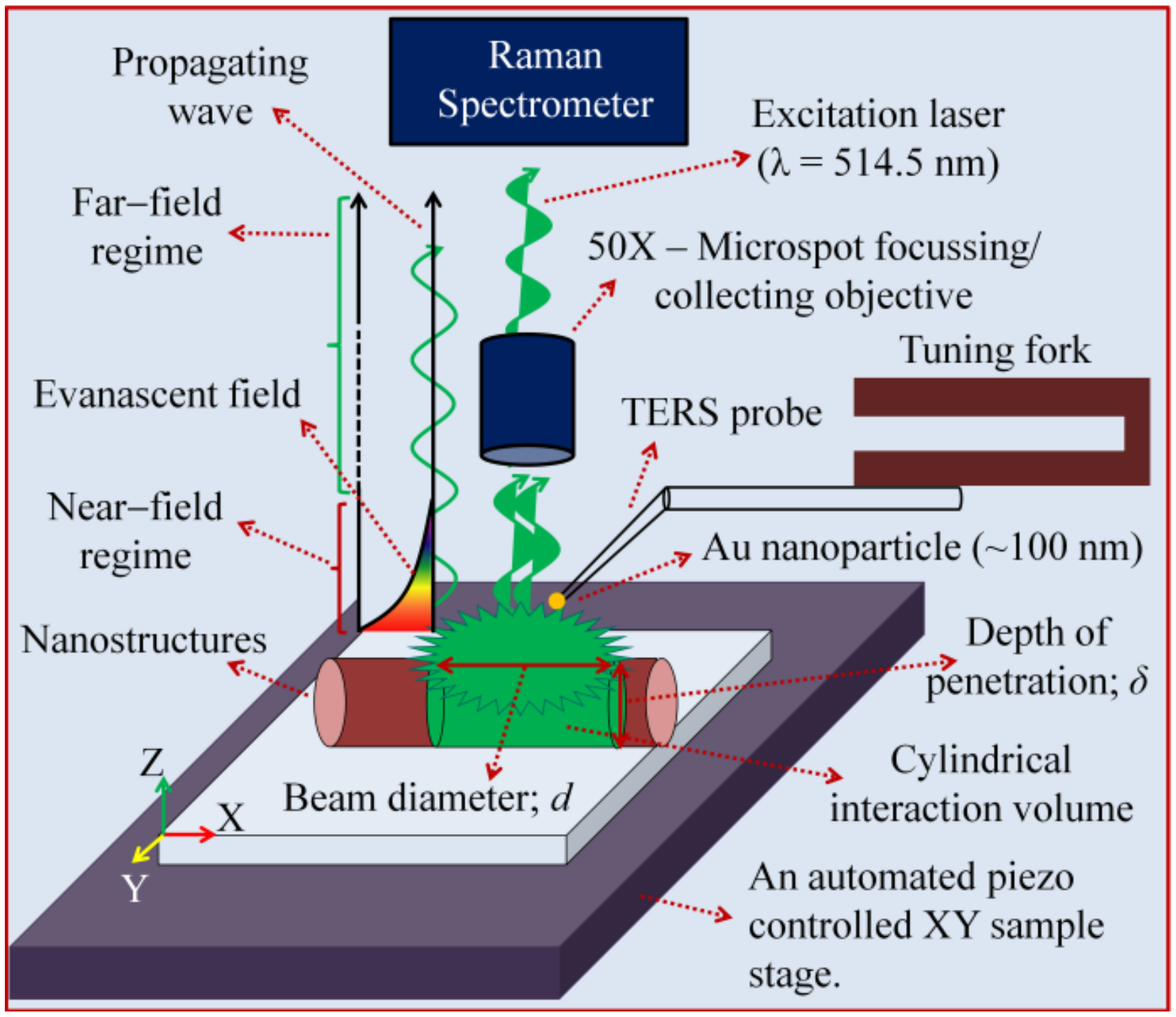

**Figure S1**. A schematic for the experimental setup of tip–enhanced Raman spectroscopy and the light–matter interaction volume generated in the nanostructure under investigation.

The interaction volume $V = \pi(\delta/2)^2 d$ is calculated as the penetration depth, $\delta$ (~3 mm) and the laser beam diameter, $d$ (~ 1 μm) are >> diameter of the nanostructures (~ 50-250 nm). So, in the present study $\delta$ and $d$ are considered as diameter and length, respectively, of the exposed volume of the 1D nanostructures where light–matter interaction takes place.

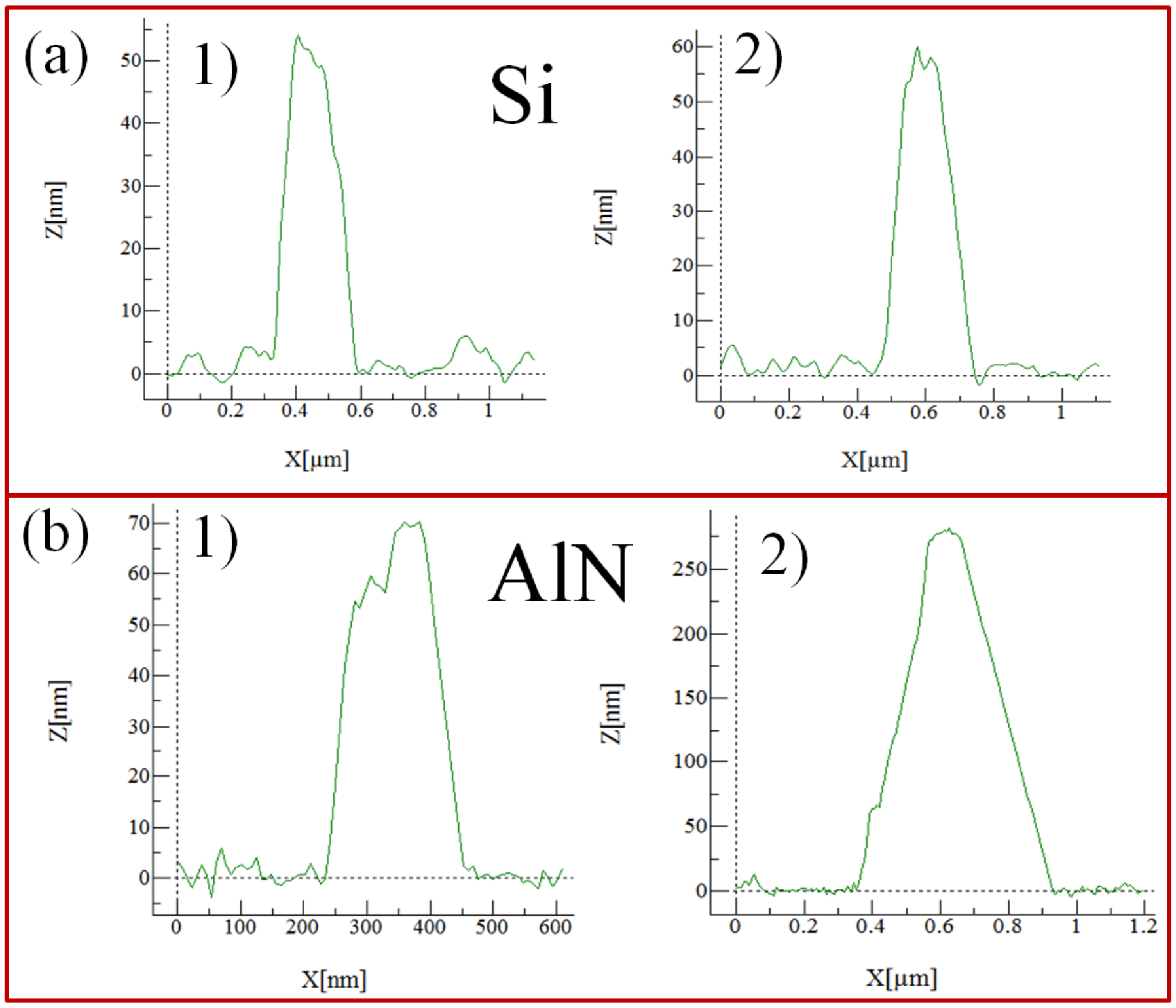

**Figure S2**. The height profiles from two different positions of the AFM topography for (a) Si NW and (b) AlN NT.